\documentclass[aps,prl,floatfix,twocolumn,superscriptaddress]{revtex4-2}
\usepackage[normalem]{ulem}
\usepackage{bm}

\usepackage{graphicx,
            caption,
            subcaption,
            amsmath,
            braket,
            units,
            ragged2e,
            xcolor,
            xspace,
            nicefrac,
            lipsum,
            nameref,
            textcomp,
            urwchancal,
            upgreek,
            isotope,
            hyperref,
            cleveref,
            siunitx,
            placeins, 
            scrextend, 
            url}

\DeclareCaptionLabelSeparator{dot}{. }
\makeatletter
\def\justified{
	\let\\\@normalcr
	\@rightskip\z@skip \rightskip\@rightskip
	\leftskip\z@skip
	\parindent 0em\relax
	\setlength{\parfillskip}{0pt plus 1fil}}
\DeclareCaptionJustification{justified}{\justified}
\hypersetup{colorlinks=true, linkcolor=blue,citecolor=blue,urlcolor=blue}
\def\unit #1 #2 {\SI{#1}{#2}\xspace}
\def\range #1 #2 #3 {\SIrange{#1}{#2}{#3}\xspace}
\DeclareSIUnit\gauss{G}

\newcommand{\myref}[2][]{Fig.~\hyperref[#2]{\ref*{#2}#1}}
\newcommand{\Myref}[2][]{Figure~\hyperref[#2]{\ref*{#2}#1}}
\newcommand{\Mytabref}[2][]{Table~\hyperref[#2]{\ref*{#2}#1}}

\begin{document}

\title{New opportunites for interactions and control with ultracold lanthanides}
\author{Matthew A. Norcia}
 \affiliation{
     Institut f\"{u}r Quantenoptik und Quanteninformation, \"Osterreichische Akademie der Wissenschaften, Innsbruck, Austria
 }
 \author{Francesca Ferlaino}
 \thanks{Correspondence should be addressed to \mbox{\url{Francesca.Ferlaino@uibk.ac.at}}}
 \affiliation{
     Institut f\"{u}r Quantenoptik und Quanteninformation, \"Osterreichische Akademie der Wissenschaften, Innsbruck, Austria
 }
 \affiliation{
     Institut f\"{u}r Experimentalphysik, Universit\"{a}t Innsbruck, Austria
 }
\begin{abstract}
Lanthanide atoms have an unusual electron configuration, with a partially filled shell of $f$ orbitals.  This leads to a set of characteristic properties that enable enhanced control over ultracold atoms and their interactions: large numbers of optical transitions with widely varying wavelengths and transition strengths, anisotropic interaction properties between atoms and with light, and a large magnetic moment and spin space present in the ground state.  These features in turn enable applications ranging from narrow-line laser cooling and spin manipulation to evaporative cooling through universal dipolar scattering, to the observation of a rotonic dispersion relation, self-bound liquid-like droplets stabilized by quantum fluctuations, and supersolid states.  In this short review, we describe how the unusual level structure of lanthanide atoms leads to these key features, and provide a brief and necessarily partial overview of experimental progress in this rapidly developing field.  
\end{abstract}

\date{March 2021}

\maketitle

\section{Introduction}

In the quest for greater control of ultracold atoms and molecules, interactions are key.  Not only do they yield interesting states distinct from their single-particle counterparts, but they also enable the evaporative cooling techniques that allow access to ultracold temperatures.  The types of interactions accessible depends strongly on the atomic species being used.  Commonly used alkali and alkaline-earth atoms interact via contact interactions, often tuned by Feshbach resonance \cite{Chin2010fri, Bloch2008mbp}, through excitation to Rydberg states \cite{gallagher2005rydberg}, or through the exchange of microwave \cite{Raimond2001} or optical photons \cite{kimble1998strong}.  Lanthanide atoms open up new opportunities for interaction, in particular due to their large ground-state magnetic dipole moments, which provide a long-range and anisotropic interaction.  Lanthanide atoms also possess a large number of optical transitions with widely varying properties, providing opportunities for control that can help exploit these interactions.

These distinctive properties originate primarily from a peculiar  electronic structure --- a so-called submerged $f$-shell configuration.  All lanthanides have a completely filled $6s$ shell, and an inner $4f$ shell that is filled to varying degree.  
Interestingly, among the different lanthanide atoms, with different numbers of valence electrons, many share a common set of properties and often feature analogous transitions at similar wavelengths, and so can in some ways be discussed interchangably.  This is in stark contrast to more commonly used species such as alkali and alkaline-earth atoms, where the addition of a single proton and electron dramatically changes the relevant level structure.

This short review provides an overview of the consequences of the unusual electron structure of the lanthanide atoms for ultracold gases. A particular focus is placed on the role of this structure in determining the optical and interaction properties of the lanthanides, and the recent experimental capabilities and observations that these have enabled.  
More detailed descriptions of the lanthanides' short-range interaction properties \cite{Kotochigova2014cib, tiesinga2021relativistic}, quantum droplets \cite{Guoreview}, and dipolar quantum gases \cite{Lahaye_2009}  can be found elsewhere.

We begin with a discussion of the complex energy level structure created by the presence of multiple valence electrons.  Next, we discuss the consequences of the large angular momentum of the ground state: the large magnetic moment, as well as the large number of accessible levels, and the resulting consequences for interparticle scattering.  Finally, as a particular direction enabled by these effects, we focus on a set of related consequences of dipolar interactions in Bose Eistein condensates (BECs): the presence of a roton excitation mode, the formation of self-bound droplets stabilized by quantum fluctuations, and the demonstration of supersolid states featuring both density modulation and global phase coherence.  We note that while ytterbium is a member of the lanthanide family, and has enabled many important works, its completely filled $f$-shell means that its properties are more akin to alkaline earth atoms, and so will not be a subject of this review.

\begin{figure}[!htb]
\includegraphics[width=3.5in, ]{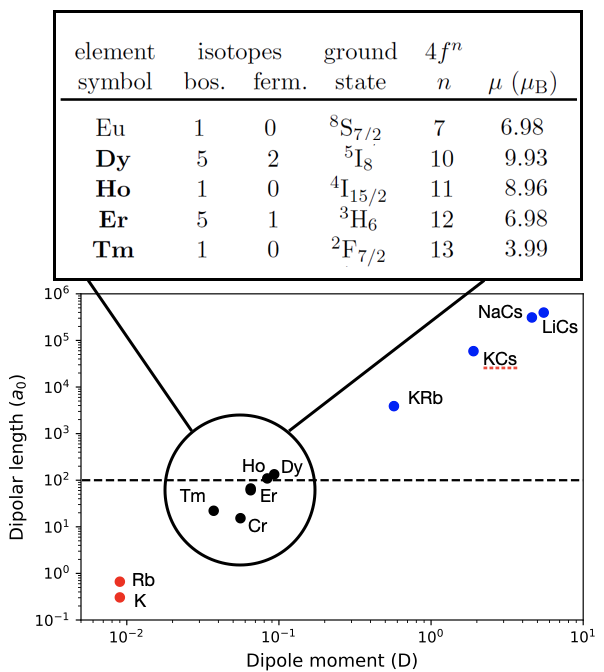}
\caption{Magnetic moments of lanthanide atoms.  Upper: key properties of the subset of lanthanide atoms that have been laser cooled, highlighting large magnetic moments and abundance of available isotopes in certain species.  Lower: dipolar length $a_{dd}$ (defined in text) versus dipole moment for a selection of alkalis, lanthanides, heteronuclear molecules, and Chromium.  Horizontal dashed line denotes 100~$a_0$, a typical scale for the s-wave scattering length.  Importantly, many lanthanides have $a_{dd}\simeq a_s$, allowing for tuning between contact-interaction dominated and dipolar-interaction dominated regimes.  }
\label{fig:introfig}
\end{figure}

\section{Complex energy spectrum}
A first important consequence of the multiple valence electrons present in lanthanides is an extraordinarily rich excitation spectrum. From ultra-narrow optical transitions in the near-infrared (few Hz) to broad transitions in the near-ultraviolet (tens of MHz),  lanthanides offer many possibilities for laser cooling, optical manipulation, and Rydberg excitations.  Such a wide range in the light-atom-coupling strength can also be found in the subset of electric dipole transitions (E2) for which the total angular momentum quantum number can vary at most by one unit ($\Delta J=0,\pm 1$). 

The gross structure of lanthanides already exhibits a variety of atomic lines. This is due both to the fact that any one of the many valence electrons can be excited, and to the strong orbital anisotropy of the electronic shells (i.e., large $L$). Considering the fine structure, the spectrum is further enriched by the spitting of energy levels and also by complex spin-orbit-coupling schemes, which can change from simple $LS$-coupling to e.g. $JJ$-coupling depending on how the angular moments of the electrons in the various sub-shells interact with each other \cite{wybourne2007optical}. 

Unlike most ultracold atoms, many lanthanides exhibit fine structure splitting even in their ground state. A recent example, attracting much attention for its application to quantum metrology, is the near infra-red inner-shell clock transition in Thulium (Tm), which connects two fine-structure levels ($J=7/2 \rightarrow J'=5/2$) of the electronic ground-state \cite{Golovizin2019}. 

In addition, lanthanides offer a wide range of clock-type transitions from the ground- to an excited electronic level. Of particular interest are the inner-shell $f-d$ orbital transitions that couple the even-parity (para) ground state to the first odd-parity (ortho) excited-level manifolds. Such transitions have Hz-range linewidth and, importantly, a wavelength in the technologically useful near-infrared range. 
Such transitions have recently been observed in dysprosium ($\lambda=1001$nm) \cite{Petersen2020} and erbium ($\lambda=1299$nm) \cite{patscheider2021observation}. For erbium, a near-magic-wavelength condition and quantum coherent manipulation has been also demonstrated.

\subsection{Laser cooling in open-shell lanthanides}

As in other multi-electron species (e.g. alkaline earth atoms), the rich internal structure of open-shell lanthanides with many $\Delta J=+1$ optical transitions in the visible and near UV offers various possibilities for laser cooling~\cite{Ban2005}. Indeed, because an important subset of the lanthanide transitions --- those involving the $s$-shell electrons --- also exist in alkaline earth atoms, many of the cooling techniques employed in lanthanides have been inspired by work in alkaline earth atoms \cite{Honda1999magneto, Katori1999magneto, Kuwamoto1999magneto}.   Moreover, because of their similarity in the atomic spectra, cooling concepts can be easily generalized between lanthanide species.
To date, laser cooling has been demonstrated in erbium~\cite{McClelland2006lcw,Berglund2007sdl,Frisch2012nlm,Seo2020epo}, dysprosium~\cite{Leefer2010tlc, Youn2010dmo, Lu2010tud,Lunden2020etc}, holmium~\cite{Miao2014mot}, thulium~\cite{Sukachev2010mot, vishnyakova2014two}, europium~\cite{Inoue2018mot}, and also in erbium-dysprosium mixtures~\cite{Ilzhofer2018tsf}. 
Further evaporative cooling has enabled the creation of degenerate gases of bosonic dysprosium \cite{Lu2011strongly} and erbium \cite{aikawa2012bose}, followed shortly by the corresponding quantum degenerate Fermi gases \cite{Lu2012quantum, aikawa2014reaching}. More recently, bosonic thulium has also been cooled to degeneracy \cite{davletov2020machine}, and quantum-degenerate lanthanide-lanthanide and lanthanide-alkali mixtures have been created \cite{Trautmann2018, Ravensbergen2018prod}.  Key properties of these laser-cooled species are provided in Fig.~\ref{fig:introfig}.

The first magneto-optical trap (MOT) of open-shell lanthanides has been realized with erbium  using a broad transition in the {\em blue} ($\lambda_{\rm Er}=401$~nm) \cite{McClelland2006lcw}. Similar transitions have been used to realize {\em blue} MOTs for dysprosium  ($\lambda_{\rm Dy}=421$~nm)~\cite{Lu2010tud}, holmium ($\lambda_{\rm Ho}=410$~nm)~\cite{Miao2014mot}, thulium ($\lambda_{\rm Tm}=410$~nm)~\cite{Sukachev2010mot}, and europium ($\lambda_{\rm Eu}=460$~nm)~\cite{Inoue2018mot}.
However, the strong radiation pressure in these blue MOTs implies a high Doppler temperature of several hundred microkelvin, so additional or alternative cooling schemes have therefore been employed to prepare samples cold enough to be loaded directly into optical traps (we note that confinement in magnetic traps is not viable because of fast spin relaxation in lanthanides~\cite{Connolly2010lsr}).
One approach has been to combine blue MOTs with an additional cooling stage, based on  a {\em red} MOT operating on a very narrow transition (from about 1 to 10 ~kHz depending on the species). Cooling into red MOTs yields temperatures as low as few $\mu$K in a spin un-polarized sample. As a second cooling stage after a blue MOT, red MOTs have been realized for erbium~\cite{Berglund2008nlm} and dysprosium~\cite{Lu2011soa}.

A different approach, now widely used in experiments for its simplicity and robustness, is to create a single {\em orange} MOT operating on an intercombination-type line of roughly 100-kHz linewidth. 
This scheme, demonstrated with erbium~\cite{Frisch2012nlm} and then with dysprosium~\cite{Maier2014nlm,Dreon2017oca,Phelps2020ssp} and erbium-dysprosium mixtures~\cite{Ilzhofer2018tsf}, allows one to reach temperatures of tens of $\mu K$, which are low enough to directly load atoms in an optical dipole trap. In addition to low temperatures, orange MOTs have two additional advantages. First, without the need for any optical pumping, they provide spin-polarized atoms in the lowest Zeeman level ~\cite{Frisch2012nlm,Maier2014nlm,Dreon2017oca}. Second, they can operate with only five beams, thus leaving vertical optical access free for e.\,g.\,microscopy~\cite{Ilzhofer2018tsf}. These peculiarities emerge from the combination of the narrow linewidth of the transition and the strongly magnetic character of lanthanides.  Recently, a combination of orange and subsequent red MOTs has been realized in erbium~\cite{Phelps2020ssp}. This is an extremely promising scheme because it allows to reach temperatures below $\mu$K, and thus substantially reducing the experimental cycle time.

\section{consequences of the large ground state angular momentum}

Because electrons in the $f$ shell have a relatively high orbital angular momentum $L=3$, partial occupancy of this shell can lead to ground state configurations with large total angular momentum.  Beyond providing a large spin space with many states that can be populated, this feature leads to the large magnetic moment and highly anisotropic properties of lanthanides; see Fig.\,\ref{fig:introfig}.

\subsection{Large magnetic moment}

Perhaps the most distinctive feature of the lanthanide atoms is their large magnetic moment, which leads to strong dipolar interactions.  In order to compare between species, it is useful to define the dipolar length $a_{dd} = m \mu_0 \mu^2/12 \pi \hbar^2$, for magnetic interactions, or $a_{dd} = m  d^2/12 \pi \epsilon_0 \hbar^2$ for electric dipoles \cite{Lahaye_2009}.  This is the length-scale at which the dipole-dipole interaction energy between two particles is comparable to the kinetic energy required by the uncertainty principle when the particles are localized to this length-scale.  

The dipolar length defined in this way is useful in that it allows us to distinguish different regimes of dipolar behavior, and place lanthanide atoms with respect to other dipolar systems, as shown in Fig.~\ref{fig:introfig}.  One productive point of comparison is to the s-wave scattering length $a_s$ through the ratio $\epsilon_{dd} = a_{dd}/a_s$.  $\epsilon_{dd} = 1$ represents the point of dipolar collapse in a homogeneous bosonic system treated at the mean-field level, and thus provides a useful boundary for strongly dipolar quantum gases \cite{Lahaye_2009}.  For typical atomic scattering lengths of around 100~$a_0$, alkali atoms (and to an even larger degree alkaline-earth atoms) are far from the strongly dipolar regime with $\epsilon_{dd} \simeq 0.01$.

First experiments on magnetic atoms focused on chromium atoms \cite{Griesmaier2005bose}, which features a large magnetic dipole moment of 6~$\mu_B$, corresponding to $\epsilon_{dd} \simeq 0.16$.  These studies enabled key observations such as dipolar collapse \cite{Lahaye2008d}, and have been reviewed in detail elsewhere \cite{Lahaye_2009}.  However,  $\epsilon_{dd}$ for chromium is small enough (due largely to its relatively small mass) that a strongly dipolar regime could only be reached by careful suppression of contact interactions near Feshbach resonances \cite{lahaye2007strong}, preventing the study of long-lived dipolar effects due to three-body losses.  
In contrast, erbium and dysprosium have dipolar lengths near 65.5~$a_0$ and 133~$a_0$ (depending on the isotope), allowing convenient access to the crossover between the contact interaction- and dipolar-dominated regimes.  Because many interesting phenomena occur near boundaries, it is actually a significant advantage for many applications to have $\epsilon_{dd} \simeq 1$, rather than $\epsilon_{dd} >> 1$.  

A particularly important implication of the long-range and isotropic nature of the dipolar interactions relates to the evaporative cooling of fermions.  Because of antisymmetrization requirements, the contact interactions that typically enable rethermalization are supressed for identical fermions at ultracold temperatures \cite{Giorgini2008theory}.  To overcome this limitation, most experimental implementations of evaporative cooling in Fermions have relied on collisions with a second species \cite{ketterle2008making} (either a different spin state, isotope, or element). 

Strong dipolar interactions provide an alternative path.  
Unlike contact interactions, dipolar interactions contribute a scattering cross-section that approaches a finite constant value at low temperatures, a phenomena referred to as universal dipolar scattering  \cite{Baranov2008tpi,bohn2009quasi}.  This makes it possible to perform effective evaporative cooling of indistinguishable dipolar Fermions all the way to the low temperatures required to reach a quantum degenerate regime.  This has been demonstrated for dysprosium \cite{Lu2012quantum} and erbium atoms \cite{aikawa2014reaching}.  For erbium, a procedure very similar to that used for Bosons could also be applied to Fermions, reaching temperatures as low as $T/T_F = 0.11(1)$, where $T_F$ is the Fermi temperature.  This low temperature enabled the observation of a deformed Fermi surface, a many-body consequence of the anisotropic interactions \cite{aikawa2014observation}.  More recently, dipolar scattering has enabled the evaporative cooling of KRb molecules \cite{valtolina2020dipolar}, highlighting the generality of this technique.  

The long-range nature of the DDI also has important consequences for integrability in one-dimensional systems --- while two-body collisions in a one-dimensional system simply exchange the momentum of two particles, leaving the overall distribution unchanged \cite{kinoshita2006quantum}, effective three-body collisions enabled by long-range interactions can redistribute momentum among three particles, leading to tunable thermalization \cite{Tang2018Thermalization}.  Further, it has recently been observed that repulsive DDI can stabilize prethermal excited states against collapse \cite{kao2021topological}.

In a regime where atoms are localized on a lattice, the dipolar interactions also provide a channel for neighboring atoms to interact, even without the possibility of occupying the same lattice site.  This situation is described in terms of extended Hubbard models, which are expected to feature density-modulated ground states with fractional filling and striped, checkerboard, or other geometric patterns \cite{baranov2012condensed, dutta2015non}.   Key features of such a model, including the presence of nearest-neighbor interactions, have been demonstrated in a bosonic erbium quantum gas confined within an optical lattice  \cite{baier2016extended}.  It is currently an active challenge to realize such a model in the context of a dipolar quantum gas microscope, where the spatial signatures of such phases could be observed at the microscopic level.

\subsection{Orbital anisotropy: dense Feshbach spectra and tensorial polarizability}

The combination of highly anisotropic orbitals and the large number of spin states leads to complex scattering interactions between lanthanide atoms.  For ultracold atoms, such scattering interactions have contributions from short-range contact interactions and longer-range dipole-dipole interactions (DDI).

The strength of the contact interaction can be tuned over a wide range by using the so-called magnetic Feshbach resonances \cite{Chin2010fri}.  The very existence, the spectral density, and the nature of such resonances largely depends on the atomic species and the interaction properties.  In alkali atoms, which have only one electron vacancy in their $s$ orbital, the scattering bond is isotropic and so is the van-der-Waals interaction, quantified by the $C_6$ dispersion coefficient, which is independent from the collision angles.  Alkali Feshbach spectra typically exhibit sparse and well isolated resonances (about one resonance per 10G or less). These resonances have been used to create tunable quantum gases and observe a variety of fascinating quantum phenomena \cite{Bloch2008mbp}. 
Alkaline-earth-like atoms, instead, have no electron vacancy in their electronic ground-state, as the $s$ orbital is fully occupied. These species have a zero electronic angular momentum and they are not expected to exhibit {\em standard} Feshbach resonances. However, we note that, by coupling two different electronic states, a new type of orbital Feshbach resonances has been recently discovered \cite{Zhang2015,Hoefer2015,pagano2015}.

\begin{figure}[!htb]
\includegraphics[width=3.5 in, ]{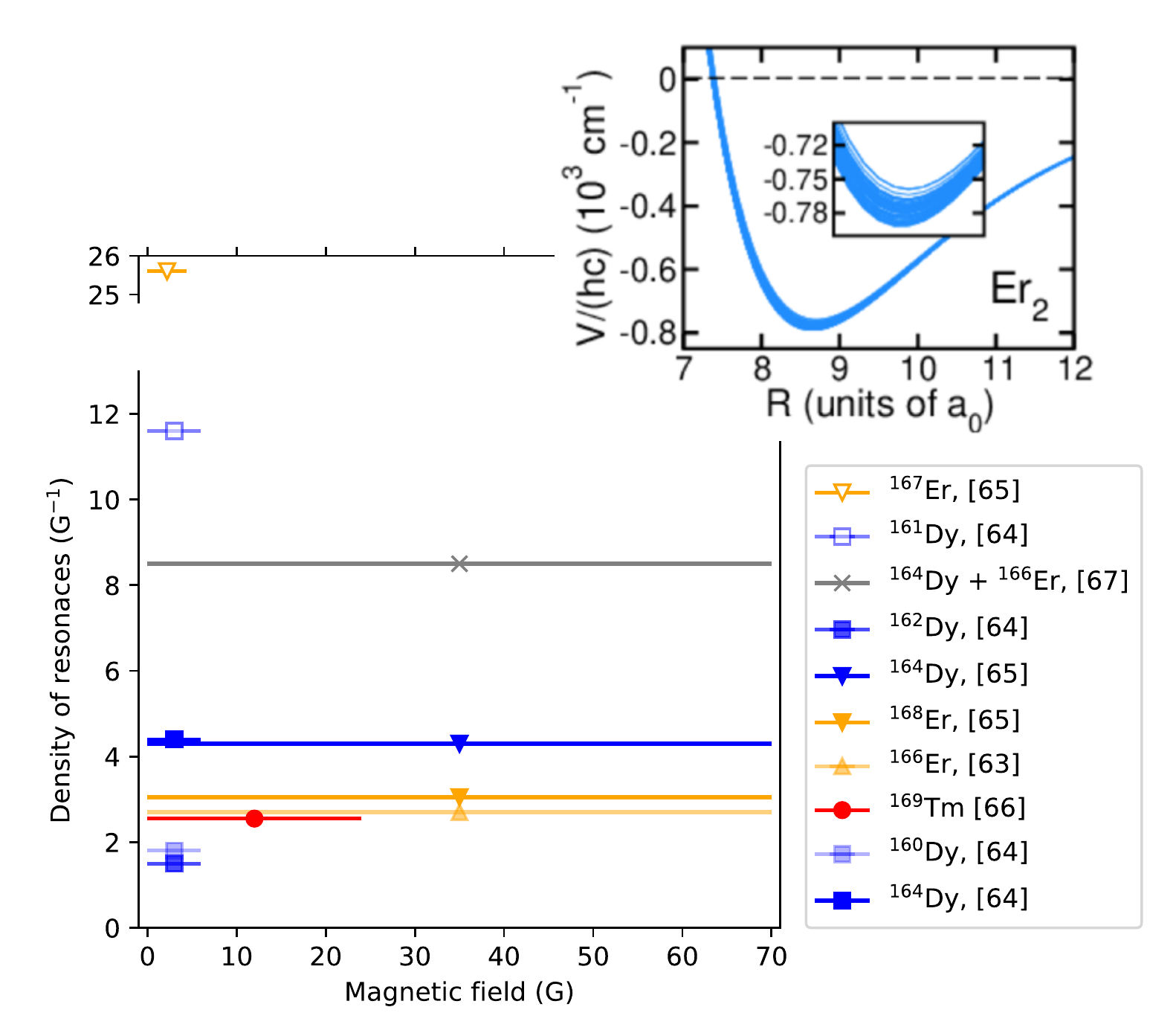}
\caption{Observed density of Feshbach resonances in lanthanide atoms.  Horizontal bars represent the magnetic field range for each measurement.  For display clarity, measurements taken at different temperatures have been averaged, and measurements of the same isotopes by the same group, providing similar results, have been omitted.  Inset: the 49 gerade Born-Oppenheimer adiabatic potentials of Er$_2$ each support bound states and contribute to the large density of Feshbach resonances.  Inset figure adapted from \cite{Li2018Orbital}. }
\label{fig:frs}
\end{figure}

The situation is rather different in open-shell lanthanides because of their orbital anisotropy. The underlying intuitive picture is simple: 
If two spheres collide, their angle of incidence does not matter (e.g. alkali). On the contrary, if two non-spherical objects collide, this angle and their shape become very important, determining the kinematics and energy associated with the collision. Quantum-mechanically, these energies are quantized and give rise to a set of many possible Born-Oppenheimer (BO) molecular potentials \cite{Kotochigova2014cib}, illustrated in Fig.~\ref{fig:frs}.  These BO potentials are comparatively close in energies, each characterized by a different $C_6$, which now depend on the magnetic quantum numbers, $C_6(m_1, m_2, m'_1,m'_2)$ \cite{Kotochigova2014cib}. Moreover, the BO potentials couple with each other, leading to an extremely large anisotropy of the interactions \cite{Petrov2012aif}. 

The many coupled BO potentials and the interaction anisotropy are at the origin of the unprecedented large number of Feshbach resonances in open-shell lanthanides, with a density several orders of magnitude higher than alkalis \cite{Kotochigova2014cib}. 
Dense Feshbach spectra have been first observed in erbium \cite{aikawa2012bose,frisch2014quantum} and then in dysprosium \cite{Baumann2014ool,Maier2015eoc},  thulium \cite{Khlebnikov2019rtc}, and erbium-dysprosium mixtures \cite{durastante2020fesh}, confirming that the high spectral density of the Feshbach resonances is a characteristic feature of open-shell lanthanides \cite{Petrov2012aif}.  Measurements of the spectral density in these various species and mixtures are summarized in Fig.~\ref{fig:frs}. 
In the low-magnetic-field region, the resonances, although typically narrow, have been successfully used to create Feshbach molecules \cite{Frisch2015ultra} and tunable dipolar quantum gases; see later discussion. Comparatively broad resonances, standing out from the forest of narrow ones, have been observed with bosonic dysprosium \cite{Maier2015buf,Lucioni2018ddb} and fermionic erbium spin-mixtures \cite{Baier2018roa}.   
Interestingly, the statistical distribution of the spacing between observed resonances reveals a chaotic character of scattering, as first observed in erbium \cite{frisch2014quantum} and then confirmed in dysprosium \cite{Maier2015eoc}, and thulium \cite{Khlebnikov2019rtc}.

The anisotropic character of the magnetic lanthanides also manifests in the atom-light interaction via the atomic polarizability, $\alpha$. This is not surprising since there is a direct relationship between the $C_6$ dispersion coefficients and the atomic polarizabilities, both being dependent on the transition dipole moments; see e.\,g.\,\cite{Kotochigova_2010}.
In the absence of hyperfine interactions, the tensor polarizability is finite only for $J > 1/2$, as is the case in the ground state of lanthanides.  We note that alkali atoms have $J = 1/2$ in the ground state, so the tensor shift is only due to hyperfine interactions, and thus falls off as $1/\Delta^2$ for $\Delta$ large compared to the hyperfine splitting, compared to $1/\Delta$ for scalar and vector components \cite{Deutsch1998quantum}.  In contrast, for states with $J > 1/2$, the tensor shift falls off instead as $1/\Delta$.  This leads to a significant tensor polarizability at large detunings, where it can be made large compared to scattering.   This opens new avenues for spin dependent trapping and manipulation, as well for a geometrical tuning of the atom-light-interaction strength. 

Because of the multi-valence-electron nature of lanthanides it is challenging to acquire an accurate knowledge of dipole polarizability. However, in the last years, tremendous progress has been made in calculating the atomic polarizability of various lanthanides; see e.\,g.\,\cite{Dzuba2011dpa,Lepers2014aot,Li2017oto,Li2017aot,Golovizin2017mfd}.
In experiments, the scalar, vectorial, and tensorial dynamical polarizability at specific wavelengths have been measured for erbium~\cite{Becher2017apo}, dysprosium~\cite{Kao2017ado, Ravensbergen2018ado,Chalopin2018als,kreyer2021measurement}, and thulium~\cite{Golovizin2017mfd, Tsyganok2019stv}. The anisotropic polarizability and spin-dependent atom-light interaction have been then used  to realize \textit{``tune-out''} ~\cite{Kao2017ado} and \textit{``magic} ~\cite{Chalopin2018als, patscheider2021observation} wavelength conditions between ground and excited states, as well as techniques for control of spin and motional dynamics, discussed in the following section.

\subsection{Large spin space}

\begin{figure}[!htb]
\includegraphics[width=3.5 in, ]{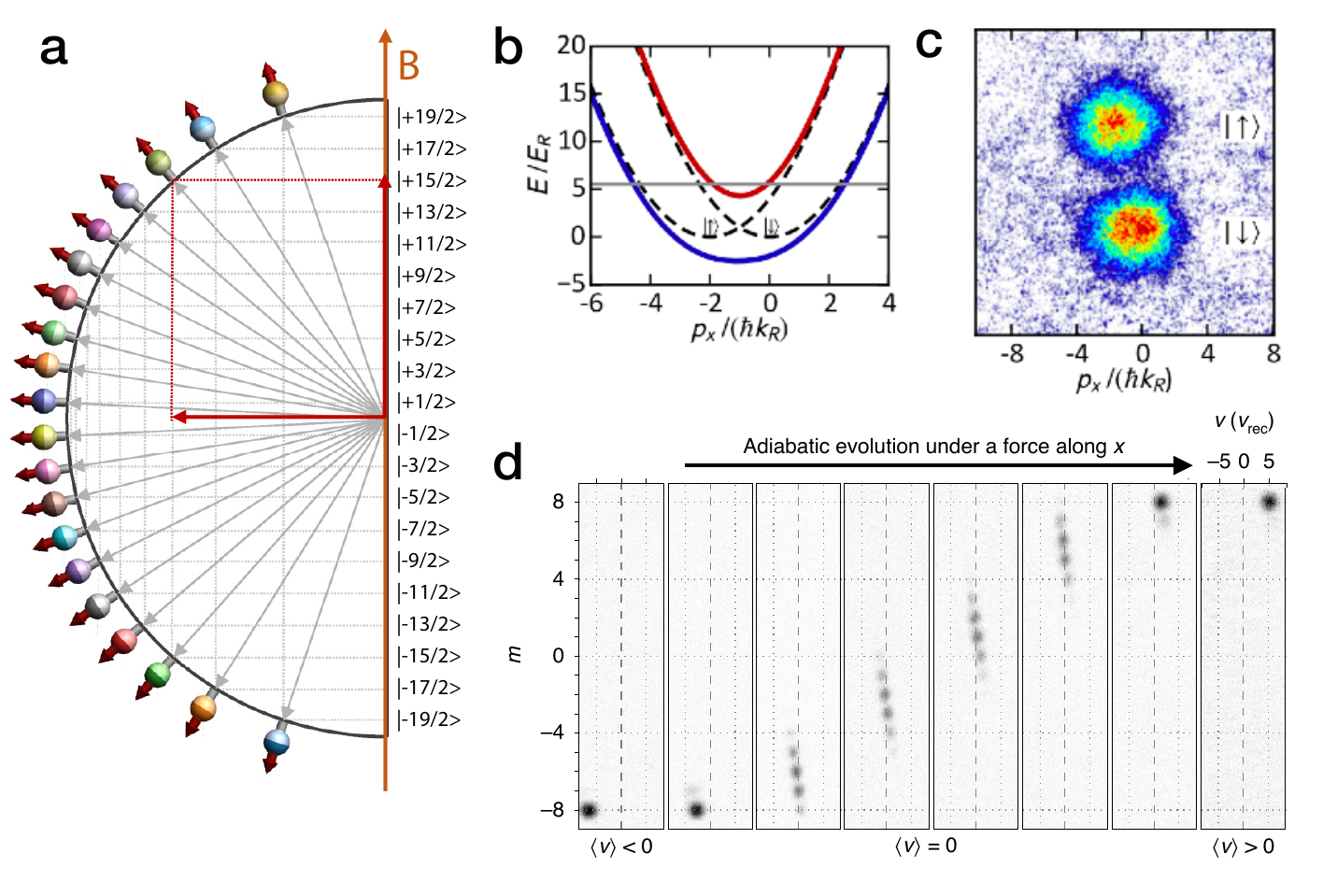}

\caption{Control over large spin space.  a) Illustration of the large spin space present in the $F = 19/2$ manifold of erbium.  Labels represent $m_f$, the projection of the total spin onto the quantization axis set by a magnetic field.  Figure reproduced from \cite{Patscheider2020controlling}.  b) Long-lived spin orbit coupling in $^{161}$Dy, enabled by large tensor light shift.  The two lowest-energy spin states ($F = 21/2$, $m_f = -21/2, 19/2$ ($\ket{\downarrow}, \ket{\uparrow}$)) are coupled by a Raman transition near $\lambda = 741$~nm, leading to a difference in momentum for the two spin states in the coupled ground-state (b), visible as the horizontal offset of the two imaged spin states in (c). Figure adapted from \cite{burdick2016long}. d)  Atomic motion within hybrid real-synthetic space created through Raman coupling among the 17 Zeeman sublevels of the $^{162}$Dy ground state.  In response to external force, the system exhibits free motion in opposite directions for the two extremal spin projections, representing edge modes, and supressed velocity in real space combined with hall drift along the synthetic dimension in the bulk.  Figure adapted from \cite{chalopin2020probing}.  }
\label{fig:spinspace}
\end{figure}

The distinctive ground-state properties of lanthanide atoms  -- large angular momentum, available narrow optical transitions, and vectorial light shifts that scale favorably relative to scattering -- provide unique opportunities for control of spin and motion within a large spin space, illustrated in Fig.~\ref{fig:spinspace}a.  Spin exchange dynamics in a large spin space within tightly lattice-confined atoms have been studied in bosonic chromium atoms \cite{lepoutre2019out}, and in fermionic erbium atoms \cite{Patscheider2020controlling}.  In the latter case, the large quadratic Zeeman and light shifts present in the fermionic atoms enabled extended control of state preparation and interactions.


These same distinctive properties have enabled new opportunities related to spin-orbit coupling (SOC).  
In bulk systems of fermionic dysprosium, the flexibility of optical control afforded by narrow linewidth transitions has enabled demonstration of long-lived spin-orbit coupling \cite{burdick2016long} (Fig.~\ref{fig:spinspace}b,c).  Here, due to the favorable scaling of the SOC strength relative to scattering, the lifetime of SOC was instead limited by collisions.  In thermal gases of bosonic dysprosium, the large spin space of the ground state provides a relatively large synthetic dimension for the exploration of spin orbit coupling in synthetic Hall systems, with distinct properties of the bulk and edges of the system \cite{chalopin2020probing} (Fig.~\ref{fig:spinspace}d). 

Non-linear light-spin interactions also enable the generation of non-classical spin states with high sensitivity to magnetic fields \cite{chalopin2018quantum}.  In this case, the tensor light shift creates a nonlinearity analogous to a one-axis-twisting type interaction between $2J$ spin 1/2 qubits.  More generally, such control over effective interactions within a relatively large spin space provides ample opportunity for studies of non-classical spin states within a highly controlled context \cite{satoor2021partitioning}.

\section{consequences of large magnetic moment in BEC: roton, droplets, and supersolid}
The interplay of dipolar and contact interactions can have dramatic consequences in quantum gases of bosonic dipolar atoms \cite{Lahaye_2009, Guoreview, Bottcher_2021}.  Varying the strength of these interactions can lead to the emergence of the so-called roton excitation mode \cite{Santos:2003, Wilson:2008, Bisset:2013, Chomaz:2018, Petter2019Probing, schmidt2021roton}, to liquid-like self-bound droplets \cite{kadau2016observing, Ferrier2016obs, wachtler2016filaments, Baillie2016self, wachtler2016droplets, schmitt2016self, Chomaz:2016, Biallie:2018, Bottcher2019dilute}, and finally to supersolid states featuring both density modulation and global phase coherence \cite{Gross:1957, Boninsegni:2012, Bottcher2019, Tanzi2019observation,Chomaz:2019}.  

\begin{figure*}[!htb]
\includegraphics[width=7 in, ]{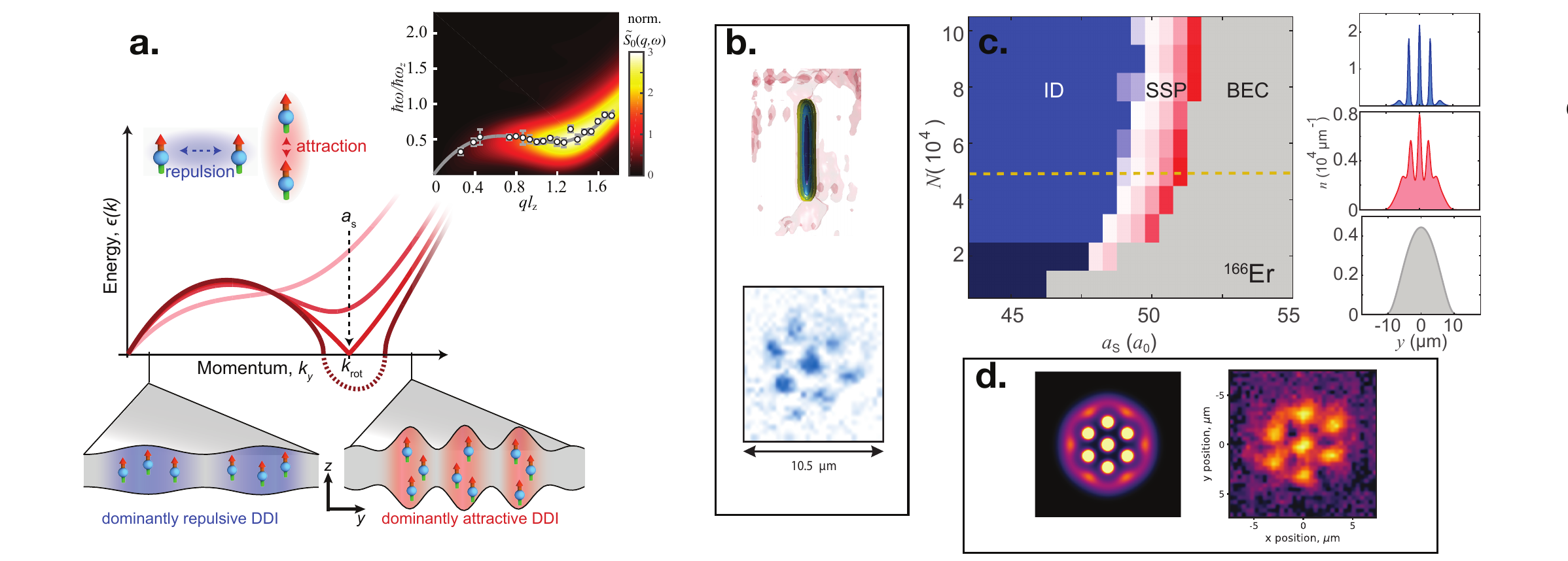}

\caption{Roton, droplets, and supersolid phase in dipolar gases.  \textbf{a.} Roton dispersion mode.  As contact interactions are reduced in a trapped dipolar quantum gas, a minimum develops in the dispersion relation, leading to modulation with a length-scale set by the confinement along the applied magnetic field.  Figure adapted from \cite{Chomaz:2018}.  Inset: roton dispersion predicted from theory and measured in experiment using Bragg spectroscopy.  Figure adapted from \cite{Petter2019Probing}.  \textbf{b.}  In the regime of dominant dipolar interactions, atoms form either a single (left) or array of (right) self-bound droplets, depending on the trap geometry.  Figures adapted from \cite{Baillie2016self, kadau2016observing}.  \textbf{c.}  Careful tuning of the contact interaction strength enables access to the supersolid phase (SSP here), between a phase of independent droplets (ID) and an unmodulated BEC.  This state features both density modulation and global phase coherence, enabled by a finite density bridge connecting high-density regions.  Figure adapted from \cite{Chomaz:2019}.  \textbf{d.} Calculated and experimentally observed density profiles for states with two-dimensional supersolidity in a nearly round trap.  Figures adapted from \cite{ss2dthy2021}.
}\label{fig:rotondropletSS}
\end{figure*}

When the strength of repulsive contact interactions dominates over that of dipolar interactions, the ground state of a dipolar condensate is a smooth, unmodulated state featuring a monotonically rising dispersion relation $\epsilon(k)$ typical of phonon excitations.  As the dipolar interaction strength increases relative to contact interactions, a minimum in the dispersion relation near $k = k_r$ begins to form, illustrated in Fig.~\ref{fig:rotondropletSS}a, indicating that density modulation at the roton wavelength $2\pi/k_r$ faces a reduced energetic penalty \cite{Santos:2003}.  This is due to the anisotropic nature of the dipole-dipole interaction --- such modulation places more atoms in a head-to-tail configuration, where the interaction is attractive, compared to in side-by-side configurations, where it is repulsive.  The degree to which atoms can be placed head to tail depends on the confinement of the trap.  This geometrical aspect of the roton mode sets the length-scale to $k_r l_z \simeq 1$, where $l_z$ is the characteristic confinement length along the magnetic field.  
If the dipole-dipole interaction becomes strong enough, $\epsilon(k_r)$ reaches zero or becomes negative, leading to the spontaneous runaway population of the roton mode, and the formation of a strongly density modulated state \cite{bohn2009does,parker2009structure}.

Such a tendency for collapse (which can be halted by the presence of beyond-mean-field effects) is a common feature of strongly dipolar gases, and manifests differently depending on trap geometry.  In particular, for anisotropic confinement that is loose along the direction of the dipoles, it leads to the formation of a single high-density region, or droplet \cite{schmitt2016self, Chomaz:2016}, (the roton wavelength in this case would be longer than the system size along directions perpendicular to the field).  When the atoms are confined within an anisotropic trap that is tighter along the magnetic field than perpendicular to it, the roton wavelength becomes smaller than the transverse size of the system.  This means that as the roton mode softens, multiple high-density regions form, eventually leading to the formation of a crystal of smaller droplets \cite{kadau2016observing, Ferrier2016obs, Wenzel2017striped}. These two situations are depicted in Fig.~\ref{fig:rotondropletSS}b. 
Crystalline structures of independent droplets have been observed in systems of dysprosium atoms confined in oblate traps \cite{kadau2016observing, Ferrier2016obs, Wenzel2017striped}. 

These droplets are of particular interest because they are not expected to be stable at the mean-field level, but rather are stabilized by beyond mean-field effects referred to as quantum fluctuations~\cite{Petrov2015qms, Ferrier2016obs, wachtler2016filaments, Baillie2016self, Chomaz:2016}.  Such stabilization was originally predicted for  mixtures of bosonic atoms with contact interactions~\cite{Petrov2015qms} (see also related mixture experiments~\cite{cabrera2018quantum, Semeghini2018sbq}), and the generality of this stabilization mechanism and its importance in dipolar droplets was soon identified~\cite{Ferrier2016obs}.
These quantum fluctuations contribute a repulsive energy penalty in the form of an effective chemical potential that scales in a local-density approximation as $n^{3/2}$, where $n$ is the local density~\cite{LHY57,Lima2011qfi}.  Because this term scales more strongly with density than the attractive mean-field interactions, for which the chemical potential scales as $n$, it prevents full collapse of the system, enabling the formation of self-bound quantum droplets.  The self-bound character of such droplets -- i.\,e.\,a dynamically stability even after external confinement is removed, originating from  a  negative  released  energy~\cite{wachtler2016filaments, Baillie2016self},  has been studied in dysprosium~\cite{schmitt2016self} and erbium~\cite{Chomaz:2016}. However, for erbium, three-body losses played a detrimental role, leading to a dynamical unbinding of the state. The observation of such droplets (as well as dipolar-stabilized excited 1D gases \cite{kao2021topological}) provided a rare example of a situation where beyond-mean-field effects lead to dramatic qualitative differences in the behavior of a quantum gas.  


The softening of the roton mode, combined with quantum stabilization, has enabled the formation of the long-sought-after supersolid state in dipolar gases \cite{Lu2015sds,Roccuzzo:2019}--- a state of matter with both periodic crystalline structure and global phase coherence\cite{Gross:1957, Andreev:1969, Chester:1970, Leggett:1970}, illustrated in Fig.~\ref{fig:rotondropletSS}c.  The presence of the low-density background indicates that atoms may move between droplets, energetically favoring configurations where all droplets share the same phase.  

Since their observation in 2019 \cite{Bottcher2019,Tanzi2019observation,Chomaz:2019}, these supersolid states have attracted intense interest and study, both theoretically in experimentally.  First experiments have been conducted in elongated, cigar-shaped traps.  In this context, studies have focused on various excitation modes of the supersolid states, which feature both superfluid and crystaline character.   In particular, the emergence of two excitation branches, each related to a spontaneous symmetry breaking, have been experimentally observed via probing symmetric axial compression modes with crystaline, superfluid, and hybrid character \cite{Natale2019, tanzi2019supersolid, Hertkorn2019fate}, as well as a low-energy Goldstone-mode \cite{guo2019low}. 
Rotational properties have been studied through measurements of angular oscillations \cite{tanzi2021evidence}, and studies of the evaporative formation and phase-locking dynamics of supersolids have explored the process by which phase coherence is established across a supersolid \cite{sohmen2021birth}.  Experiments have also demonstrated the presence of a dissipative mechanism that causes the phase of the high-density regions to synchronize and lock in the phase-coherent configuration \cite{Ilzhofer:2019}.

Following the demonstration of 1D supersolidity in dipolar gases, its two dimensional (2D) counterpart is currently attracting substantial renewed interest. 
Theoretically, recent studies have focused on the phase diagram and excitation spectra of 2D supersolid in isotropic and anisotropic traps \cite{zhang2021phases, hertkorn2021supersolidity,norcia2021two}, the possibility of such supersolids hosting vortices in the low-density regions \cite{ Roccuzzo2020rotating, Ancilotto2021vpi, Gallemi2021vortex}, and on the formation of novel type of ground-states with exotic spatial patterns \cite{ zhang2021phases, hertkorn2021pattern}. 
Very recently, supersolidity along two dimensions has been demonstrated in a trap of variable anisotropy \cite{norcia2021two}.  In a related effort, an angular roton mode, analogous to the linear roton mode present in elongated traps, has been observed in a  skittle shaped trap \cite{schmidt2021roton}. 

\section{Conclusions}
This short review has attempted to provide an overview of new opportunities for atomic interactions and control that are enabled by the unusual electronic configuration of lanthanide atoms.  Many of these directions have only recently been explored, and we expect this to represent only the beginning of the lanthanide story for ultracold quantum gases.  An extended review on dipolar gases is currently in preparation \cite{dipolarrev2021}.

\begin{acknowledgments}

\textbf{Funding:} 
F.F. is financially supported through an ERC Consolidator Grant (RARE, No.\,681432), an NFRI grant (MIRARE, No.\,\"OAW0600) of the Austrian Academy of Science, the QuantERA grant MAQS by the Austrian Science Fund FWF No\,I4391-N.  F.F. acknowledge the DFG/FWF via FOR 2247/PI2790. 
M.A.N.~has received funding as an ESQ Postdoctoral Fellow from the European Union’s Horizon 2020 research and innovation programme under the Marie Skłodowska‐Curie grant agreement No.~801110 and the Austrian Federal Ministry of Education, Science and Research (BMBWF).

\end{acknowledgments}

\bibliography{references}
\end{document}